# Title: Echography of Young Stars Reveals Their Evolution

**Authors:** K. Zwintz[1]*, L. Fossati[2], T. Ryabchikova[3], D. Guenther[4], C. Aerts[1,5], T. G. Barnes[6], N. Themeßl[7], D. Lorenz[7], C. Cameron[8], R. Kuschnig[7], S. Pollack-Drs[7], E. Moravveji[1], A. Baglin[9], J. M. Matthews[10], A. F. J. Moffat[11], E. Poretti[12], M. Rainer[12], S. M. Rucinski[13], D. Sasselov[14], W. W. Weiss[7]

**Affiliations:**

[1]Instituut voor Sterrenkunde, KU Leuven, Celestijnenlaan 200D, 3001 Leuven, Belgium.

[2]Argelander Institut für Astronomie der Universität Bonn, Auf dem Hügel 71, 53121 Bonn, Germany.

[3]Institute of Astronomy, Russian Academy of Sciences, Pyatnitskaya 48, 109017 Moscow, Russia.

[4]Department of Astronomy and Physics, St. Mary's University, Halifax, NS B3H 3C3, Canada.

[5]Department of Astrophysics, IMAPP, Radboud University Nijmegen, P.O. Box 9010, 6500 GL Nijmegen, The Netherlands.

[6]The University of Texas at Austin, McDonald Observatory, 82 Mt. Locke Rd., McDonald Observatory, Texas 79734, USA.

[7]Universität Wien, Institut für Astrophysik, Türkenschanzstraße 17, 1180 Vienna, Austria.

[8]Department of Mathematics, Physics & Geology, Cape Breton University, 1250 Grand Lake Road, Sydney, Nova Scotia, Canada, B1P 6L2.

[9]LESIA, Observatoire de Paris-Meudon, 5 place Jules Janssen, 92195, Meudon, France.

[10]Department of Physics and Astronomy, University of British Columbia, 6224 Agricultural Road, Vancouver, BC V6T 1Z1, Canada.

[11]Départment de Physique, Université de Montréal, C.P. 6128, Succ. Centre-Ville, Montréal, QC H3C 3J7, Canada.

[12]INAF-Osservatorio Astronomico di Brera, Via E. Bianchi 46, 23807 Merate, Italy.

[13]Department of Astronomy and Astrophysics, University of Toronto, 50 St. George St., Toronto, ON M5S 3H4, Canada.

[14]Harvard-Smithsonian Center for Astrophysics, 60 Garden Street, Cambridge, MA 02138, USA.

*Correspondence to: konstanze.zwintz@ster.kuleuven.be.

**Abstract**: We demonstrate that a seismic analysis of stars in their earliest evolutionary phases is a powerful method to identify young stars and distinguish their evolutionary states. The early star that is born from the gravitational collapse of a molecular cloud reaches at some point sufficient temperature, mass and luminosity to be detected. Accretion stops and the pre-main sequence star that emerges is nearly fully convective and chemically homogeneous. It will continue to contract gravitationally until the density and temperature in the core are high enough to start nuclear burning of hydrogen. We show that there is a relationship between detected pulsation properties for a sample of young stars and their evolutionary status illustrating the potential of asteroseismology for the early evolutionary phases.

**One Sentence Summary:** We interpret the pulsational properties of young stars that have not yet entered their nuclear burning phases and show that a seismic analysis can be used to determine the young stars' evolutionary states.

**Main Text:**

The earliest phases in the lives of stars determine their future fate. For example, the production of chemical elements, which are used to trace the history of Galactic evolution, depends on the initial mass and metallicity of young stars. Also, the angular momentum obtained during the birth process is vital for the further stellar life. Therefore, understanding the physical processes that occur during the earliest stages of stellar evolution is essential.

During the early evolutionary phases, following the star's appearance on the birthline (*1, 2, 3*), the central temperatures and densities are not yet high enough to initiate significant nuclear burning, hence, the early star derives all of its energy (heat and radiative) from the release of gravitational potential energy. The interior structure undergoes several changes as the temperature rises and the opacities, in general, decrease (*1*). Initially, the higher opacities and the outward transportation of the released gravitational energy force the star to be nearly fully convective. But, as the star heats up and the opacities drop, a radiative core develops. At this point the star leaves the Hayashi track (*4*), follows the Henyey track (*5*), moving leftward to hotter surface temperatures in the evolutionary diagram. Eventually, the star's central regions become hot enough for nuclear burning to occur. Models predict that for intermediate-mass stars (i.e., with 1.5 to 8 solar masses, $M_\odot$), just before the star reaches this nuclear burning stage, called the main-sequence (MS) phase, there is a "bump" or brief period of nuclear burning where carbon is "burned" by the CNO cycle (a chain of hydrogen burning reactions catalyzed by the presence of C, N, and O) until C and N reach nuclear burning equilibrium. As this happens, the star's core briefly becomes convective and the star's luminosity increases. Shortly thereafter, the star resumes its contraction onto the zero-age main sequence (ZAMS), when nuclear burning of hydrogen re-ignites and remains the dominant source of energy throughout most of the rest of the star's life. The precise definition of the ZAMS is critical when comparing ages of young stars. Here we define the ZAMS as the time following the CN equilibrium burning bump when the nuclear burning accounts for at least one percent of the total luminosity of the star.

For pre-MS stars with 1.5 to 4 $M_\odot$, the interior structure will at some point during their evolution along the Henyey track become unstable to acoustic oscillation modes, which can be observed at the surface as periodic variations in luminosity and temperature or as Doppler shifts of spectral lines. So, for at least some pre-MS stars, it is possible at some evolutionary stages to use asteroseismology to explore their structure and evolution. The pre-MS stars addressed in this study show coherent self-excited low-radial-order pressure (p) modes (*6*) with periods from ~18 minutes up to 6 hours. Until 2008, only 36 of them were known (*7*); through 2011, the number increased significantly to more than 60 (*8*) using dedicated ground-based observations and high-precision time-series photometry obtained with the Microvariability and Oscillations of STars (MOST) (*9*) and COnvection, ROtation et Transits planétaires (CoRoT) (*10*) satellites. Although the pulsations of pre-MS stars could be studied in relative detail, little information was available about the stars' fundamental astrophysical properties, such as effective temperature ($T_{\mathrm{eff}}$), surface gravity (log $g$), luminosity, mass (M), projected rotational velocity ($v\sin i$) and metallicity ([$Fe/H$]), due to the lack of spectroscopic measurements for this type of objects.

Our sample comprises 34 stars with confirmed pre-MS evolutionary status, pulsational variability and accurate fundamental parameters determined from detailed spectroscopic analyses. We assessed the pre-MS status of the stars in our sample in two different ways: through their membership to open clusters known to be younger than 10 million years, which ensures that 1.5-4 $M_\odot$ stars have not reached the ZAMS yet; or by their identification as Herbig Ae stars (*11*), which are associated with star-forming regions.

The pulsational properties of the sample stars were determined from ground and space-based observations. For nineteen stars, high-precision space photometry from the MOST (*9*) and CoRoT (*10*) satellites was used to investigate the pulsations. For eight of those (i.e., HD 261737, HD 262014, HD 262209, HD 37357, NGC 2244 45, NGC 2244 183, NGC 2244 271 and NGC 2244 399) the pulsational variability was discovered in the course of this study, while the seismic properties for the remaining eleven objects observed from space have been described in previous work (*see Supplementary Material for all references*). Additionally, the pulsational properties for 15 stars were taken from ground-based observing campaigns reported in the literature (*see Supplementary Material for all references).*

For the determination of the fundamental parameters, we obtained high-resolution, high signal-to-noise ratio spectra for 27 objects of our sample to complement the variability analysis; for the other seven stars, fundamental parameters were available in the literature *(see Supplementary Material).*

To put the stars' positions in the $T_{\mathrm{eff}}$ – log $g$ plane in context to their relative evolutionary stage, we computed pre-MS evolutionary tracks with the Yale Stellar Evolution Code (YREC, *12*) for solar metallicity (i.e., Z = 0.02) (Fig. 1, 2). We find that our sample includes stars that gain their energy solely from gravitational contraction (i.e., stars that are on the Henyey track just before arriving on the ZAMS), stars in which incomplete nuclear burnings have started (i.e., stars that are briefly burning CN producing the short rise in the luminosity before the star settles onto the ZAMS as discussed above), and stars close to the ZAMS where hydrogen-burning in the cores has already started.

In general, the stellar lifetime strongly depends on the birth mass. In the case of pre-MS stars, this dependence is manifested by the Kelvin-Helmholtz contraction time scale (*13*) rather than the nuclear time scale that rules the evolution of MS stars. The 34 pre-MS stars of our sample have masses between 1.5 and 3.0 $M_\odot$. Although this mass range is rather narrow, the differences in stellar ages at the ZAMS are significant: a star with 3.0 $M_\odot$ needs only ~8 million years from the birthline (i.e., the point where the star emerges from the molecular cloud and mass accretion stops) until hydrogen is ignited in its core (i.e., the arrival on the ZAMS), while a star with 1.5 $M_\odot$ has a ZAMS age of ~33 million years. The duration of the star formation process is thus strongly dependent on the mass of the pre-MS star.

The evolution of stellar rotation rates during the pre-MS phases is poorly understood. As young stars contract during their transition from the birthline to the ZAMS, stellar evolution theory predicts an increase in their rotational velocities if angular momentum is conserved (*14*). The fraction of the projected rotational velocity ($v\sin i$) over the critical breakup velocity ($v_{\mathrm{crit}}$) for the 34 pre-MS stars (Fig. 1) indeed indicates the occurrence of higher rotation rates at the times of the first nuclear burning compared to the one of stars in the earlier stages. We find that the majority of the pre-MS stars in our sample rotates at a low to moderate fraction of their critical breakup velocity (Fig.S3), which points to angular momentum loss by, e.g., winds or star-disk coupling. Our results agree with recent results reported for Herbig Ae/Be stars in a similar mass range (*15*).

The acoustic cut-off frequency is the highest frequency of a pressure mode that can still be trapped inside the star and is therefore an important seismic characteristic. Theoretical computations for pre-MS stars predict the highest values to occur close to the ZAMS (i.e., around 1000μHz or 86.4d$^{-1}$) and the lowest values (i.e., around 100μHz or 8.6d$^{-1}$) close to the birthline (*16*). For high-order pressure modes, the acoustic cut-off frequency scales with the frequency of maximum power, $\nu_{max}$ (*17*). For comparison with our observational sample, we used the highest significant pressure-mode frequency, $f_{max}$, which is the measurable quantity in our data coming closest to the acoustic cut-off frequency computed from seismic models. Our observations indeed confirm the theoretical predictions: the coolest and least evolved stars have the lowest $f_{max}$ values, and therefore pulsate with the longest periods (Fig. S4). The hottest and most evolved stars have the highest $f_{max}$ values and pulsate with the shortest periods. A similar relation is found when we use $\nu_{max}$ derived from our data instead of $f_{max}$. In the present sample $f_{max}$ ranges from 76.47 μHz (for the star HR 5999) to 970.37 μHz (for the star HD 261711).

Our sample consists of a mixture of different types of pre-MS objects (cluster members and Herbig Ae field stars) that also have a different formation history. By taking a subsample of pulsating stars that formed from the same molecular cloud and had the same initial conditions, we reduced possible ambiguities in the interpretation. In the young open cluster NGC 2264 (*18, 19*) nine pulsating pre-MS members were discovered by multi-epoch observations conducted with the MOST and the CoRoT space telescopes. The pulsational variability of HD 261737, HD 262014 and HD 262209 is discovered in this work (*see Supplementary Material*), while the seismic properties for V 588 Mon, V 589 Mon, HD 261230, HD 261387, NGC 2264 104 and HD 261711 have been described in previous work (*see Supplementary Material for all references*). The nine pre-MS pulsators in NGC 2264 (marked as "a" through "i" in Fig. 2 and Fig. 3) have different masses and are found to be in different relative evolutionary phases. Their evolutionary tracks in conjunction with their pulsational properties (Fig. 2 and Fig. 3) illustrate that the least-evolved stars pulsate slower (V 589 Mon) than the objects that are already located closer to the ZAMS (HD 261711). This is evidence of sequential star formation, making the assignment of a single age to this and other young open clusters unjustified. In particular, we find that the duration of the star formation process in NGC 2264 is going on already much longer than the cluster age of 1 to 5 Myr reported in the literature (e.g., *18, 19*).

A clear relation between the seismic observables and evolutionary status for the pre-MS stars emerges, even though our sample is relatively heterogeneous due to both the mixture of field and cluster stars of varying metallicity as well as different precision for the available data sets. This relation is even clearer when we consider stars formed from the same birth environment, such as members of the young cluster NGC 2264. We have shown with this study that asteroseismology can discern different stages of evolution not only for dying intermediate-mass stars as was recently found (*20*), but can be similarly applied to unravel the star formation history of young stars and even deliver the relative ages of the various mass populations of pre-MS stars of young open clusters.

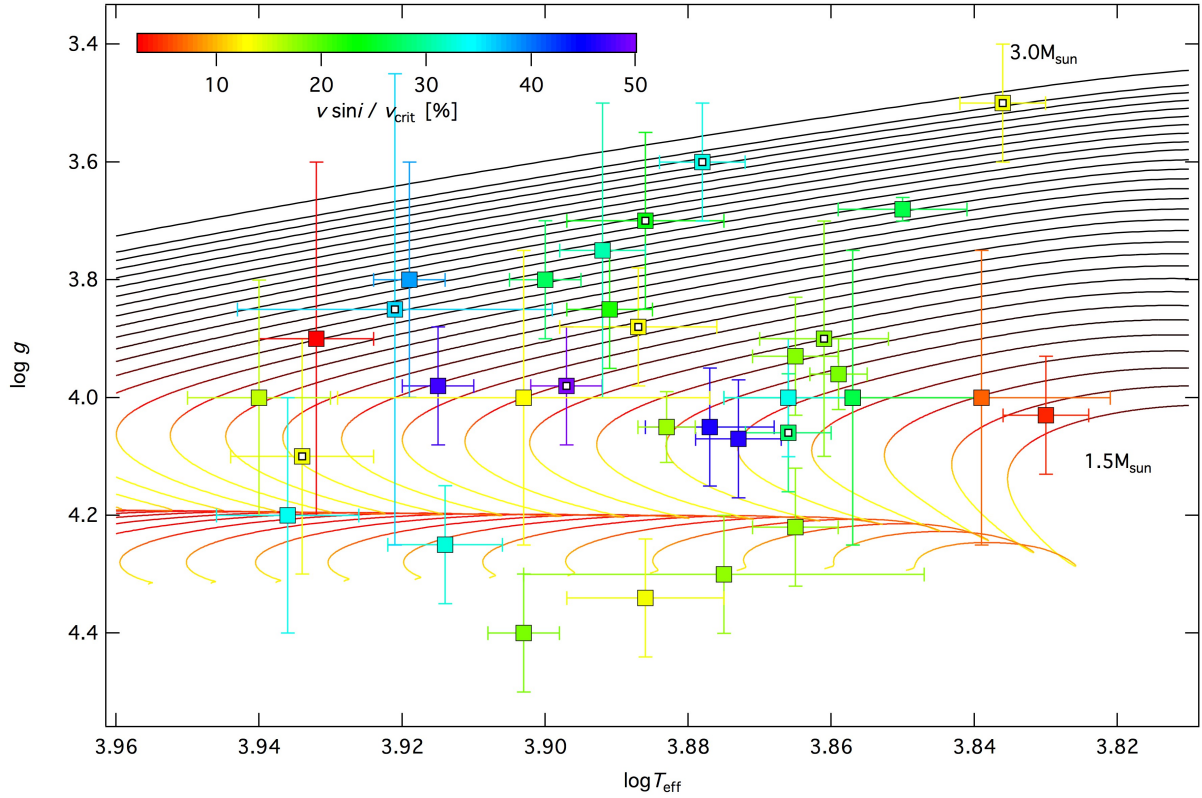

**Fig. 1**. Evolutionary diagram of surface gravity against effective temperature showing the 34 pulsating pre-MS stars. Symbol colors correspond to the percentage of critical breakup velocity ($v\sin i / v_{\mathrm{crit}}$) according to the scale at top-left. Pre-MS evolutionary tracks (*12*) are shown from 1.5 to 3.0 $M_\odot$ in steps of 0.05 $M_\odot$ and are color-coded according to the contribution of gravitational potential energy to the total energy output: when gravitational contraction is the only energy source, the pre-MS tracks are drawn in black; the track brightness increases as the ratio of gravitational potential energy to the total energy output drops, with the onset of first non-equilibrium nuclear burnings closer to the ZAMS. The symbols marked with a white smaller square in their centers are members of the young cluster NGC 2264 labeled "a" to "i" in Fig. 2 and Fig. 3.

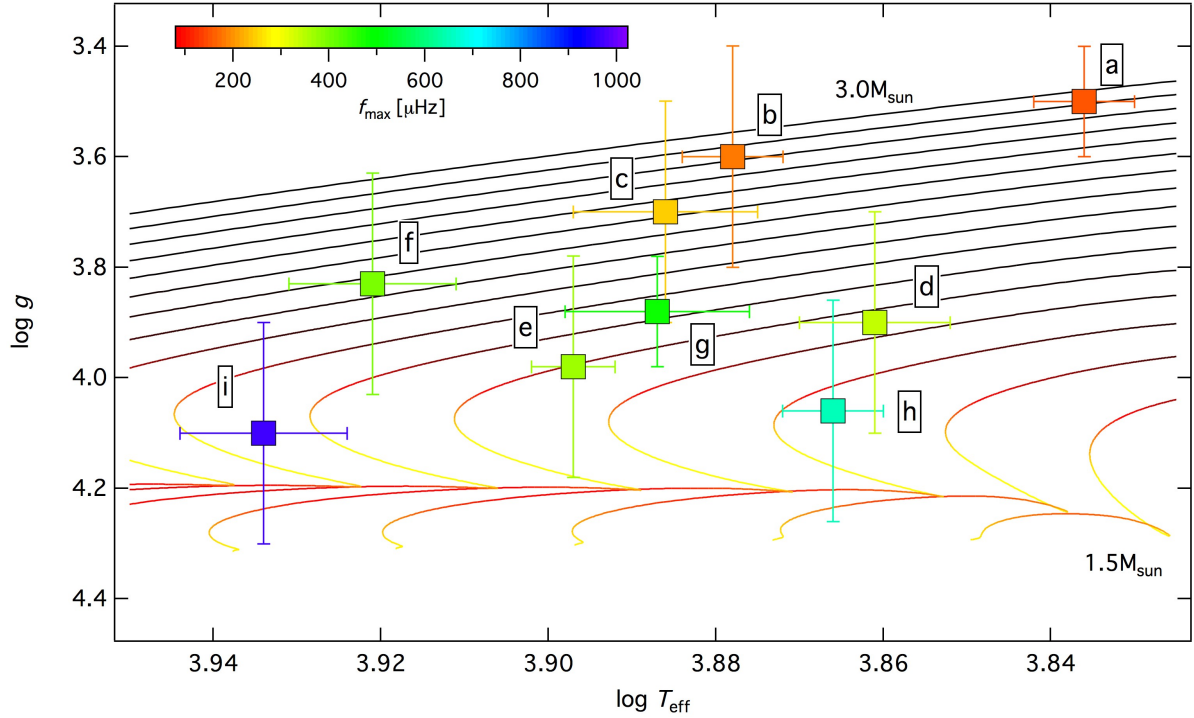

**Fig. 2**. Evolutionary diagram of surface gravity against effective temperature showing a subset of nine pulsating pre-MS stars in the young cluster NGC 2264: V 589 Mon, V 588 Mon, HD 261230, HD 262209, HD 261737, HD 261387, HD 262014, NGC 2264 104 and HD 261711 labeled "a" to "i. Symbol colors correspond to the highest observed p-mode frequency, $f_{max}$, according to the scale at top-left. Pre-MS evolutionary tracks (*12*) are shown from 1.5 to 3.0 $M_\odot$ in steps of 0.1 $M_\odot$ and are color-coded according to the contribution of gravitational potential energy to the total energy output (as in Fig. 1). The amplitude spectra of these nine objects are shown in Fig. 3 using the same labels.

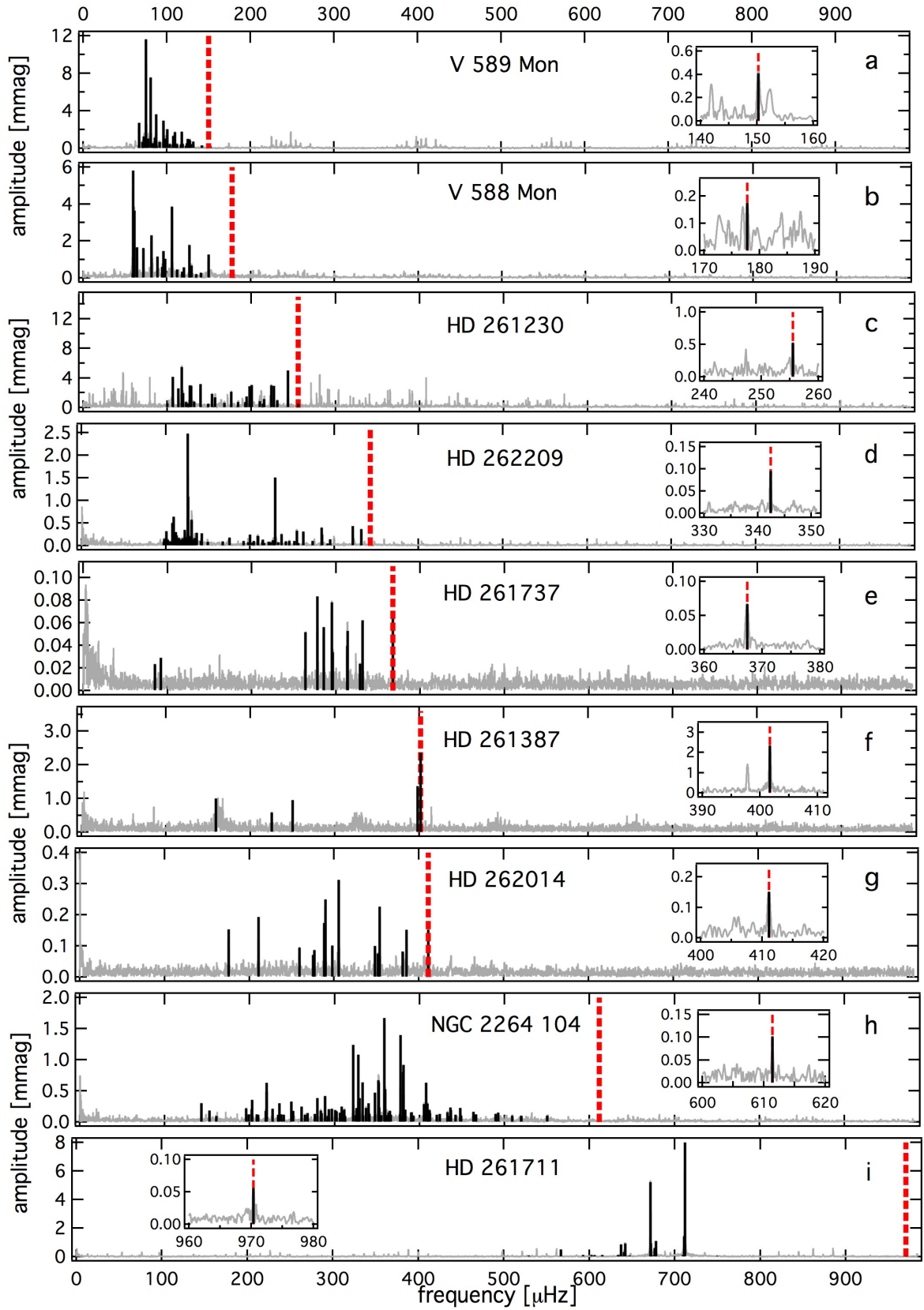

**Fig. 3.** Sequence of amplitude spectra for a subset of nine pulsating pre-MS stars in the young cluster NGC 2264 from least evolved (V 589 Mon), to most evolved (HD 261711). In each panel the X-axis shows frequency in μHz and the amplitudes on the Y-axis are given in millimagnitudes. The original amplitude spectra are shown in light grey, where all identified pulsation frequencies are marked with black solid lines. The location of the maximum pulsation frequency, $f_{max}$, for each star is marked with the dashed red line. The small insets show the region immediately around $f_{max}$ to illustrate its statistical significance.

**Acknowledgments:** Data obtained with the MOST satellite are available through the MOST Public Data Archive (http://most.astro.ubc.ca//archive.html) and data from CoRoT in the CoRoT Archive (http://idoc-corot.ias.u-psud.fr/index.jsp). Spectroscopic data can be obtained at the ESO archive (http://archive.eso.org) and the CFHT archive (http://www3.cadc-ccda.hia-iha.nrc-cnrc.gc.ca/en/cfht/). Data are also available from the public website http://www.ster.kuleuven.be/~konstanze/science/.

We are grateful to the MOST and CoRoT teams for making this work possible. We would like to thank N. Piskunov for his advice on the best usage of the SME software. The research leading to these results has received funding from the European Research Council under the European Community's Seventh Framework Programme (FP7/2007 – 2013) / ERC grant agreement No 227224 (PROSPERITY), from the Research Council of the KU Leuven under grant agreement GOA/2013/012, and from the Fund for Scientific Research of Flanders (FWO), Belgium, under grant agreement G.0B69.13.

TR acknowledges partial financial support from the Presidium RAS Program "Nonstationary Phenomena in Objects of the Universe". DBG, JM, AFJM and SMR acknowledge the funding support of the Natural Sciences and Engineering Research Council (NSERC) of Canada. RK and WWW are supported by the Austrian Fonds zur Förderung der wissenschaftlichen Forschung (P22691-N16) and by the Austrian Research Promotion Agency-ALR. EP and MR acknowledge financial support from the FP7 project "SPACEINN: Exploitation of Space Data for Innovative Helio- and Asteroseismology". EM is beneficiary of a postdoctoral grant from the Belgian Federal Science Policy Office co-funded by the Marie Curie Actions FP7-PEOPLE-COFUND2008 n246540 MOBILITY GRANT from the European Commission. CC was supported by the Canadian Space Agency.

Spectroscopic data were obtained with the 2.7-m telescope at Mc Donald Observatory, Texas, US, the 3.6m-telescope at La Silla Observatory under the ESO Large Programme LP185.D-0056, the CFHT (proposal number CFHT2012AC003), the CTIO 1.5m telescope and the Mercator Telescope (operated on the island of La Palma by the Flemish Community, at the Spanish Observatorio del Roque de los Muchachos of the Instituto de Astrofísica de Canarias) with the HERMES spectrograph, which is supported by the Funds for Scientific Research of Flanders (FWO), Belgium, the Research Council of K.U. Leuven, Belgium, the Fonds National Recherches Scientific (FNRS), Belgium, the Royal Observatory of Belgium, the Observatoire de Genève, Switzerland and the Thüringer Landessternwarte Tautenburg, Germany.

**Supplementary Materials:**

Materials and Methods

Figures S1-S6

Table S1

References (*21-68*)

**Supplementary Materials:**

**1. Photometric time series: observations, data reduction and frequency analysis**

The pulsational variability of the 34 pre-MS stars was investigated using ground and space based observations. High-precision time series photometry from the MOST space telescope was obtained for 13 stars, from the CoRoT satellite for three stars, and from both satellites, MOST and CoRoT, for four additional stars. Ground-based data were available for 22 stars; part of the information used here was taken from the literature. Table S1 lists the 34 pre-MS stars of our sample, the highest pressure-mode frequency, $f_{max}$, the data from which the pulsational variability was assessed (MOST, CoRoT, ground based) and references to the literature where appropriate.

For eight out of our sample of 34 stars, pulsational variability has been discovered in the course of this work. Three of those eight stars are members of the young cluster NGC 2264, and their amplitude spectra are shown in Fig. 2 labeled "e", "g" and "d", respectively: HD 261737 was found using MOST data, while HD 262014 and HD 262209 were observed with the CoRoT satellite. The other five new pre-MS pulsators are the Herbig Ae star HD 37357 and the four members of the young cluster NGC 2244 named NGC 2244 45, NGC 2244 183, NGC 2244 271 and NGC 2244 399. Data for these five stars were obtained with the MOST space telescope. The corresponding amplitude spectra are shown in Fig. S1.

For six out of our sample of 34 stars the presence of pulsations has been reported earlier in the literature. For the present study we used new and previously unpublished data sets of higher precision obtained by the MOST space telescope for these objects. Three of these six stars are members of the young cluster NGC 2264. Their amplitude spectra are shown in Fig. 2 labeled "c", "f" and "h". The pulsational variability of the stars HR 5999, V 1247 Ori and V 351 Ori was discovered earlier using ground-based data (*21*, *22*, *23*). The results of our pulsational analysis conducted with data obtained by the MOST space telescope are shown in Fig. S2.

We point out that uncertainties arise from the mixture of different types of data sets used to determine $f_{max}$: photometric time series obtained from space have significantly higher precision, hence lower noise levels, than most of the ground based observations. This propagates in lower seismic quality of the $f_{max}$ values for some of the stars observed with less precision.

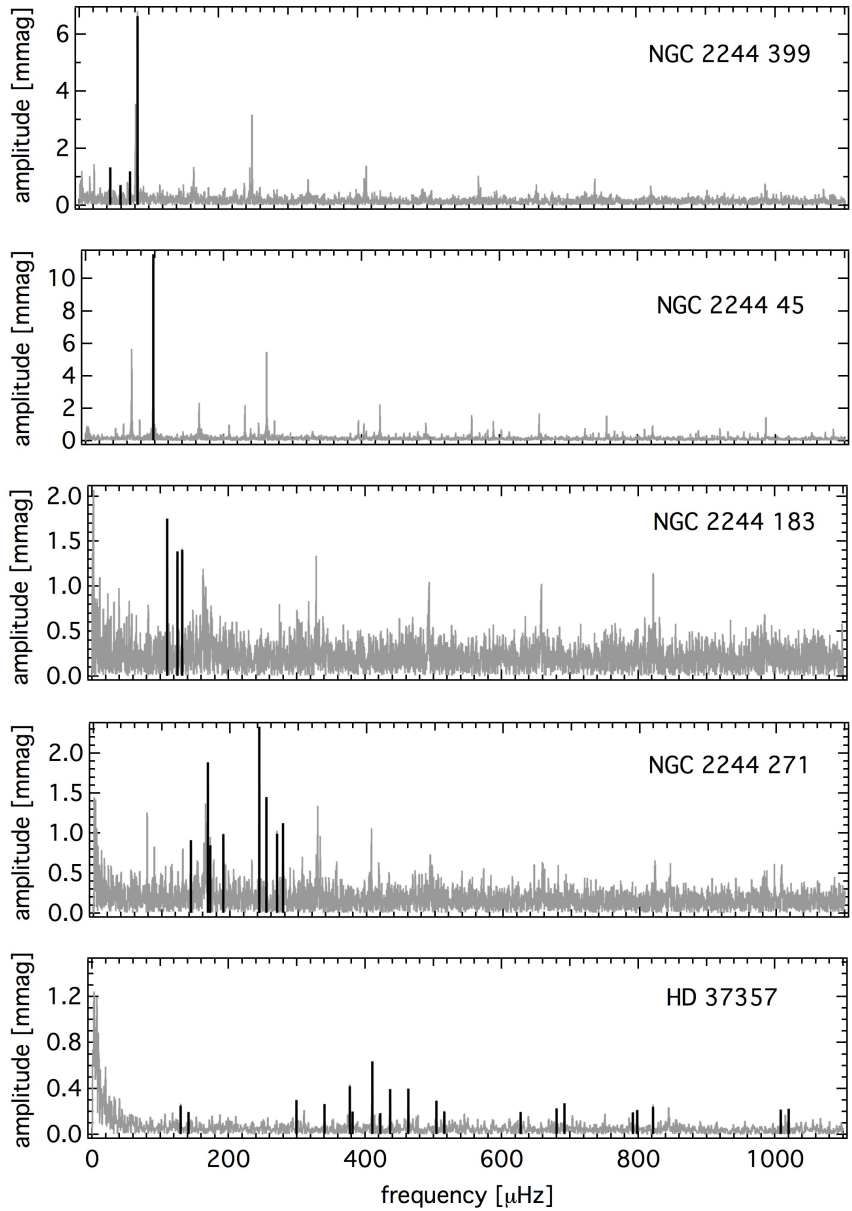

**Fig. S1.** Amplitude spectra for the five new pre-MS pulsators NGC 2244 399, NGC 2244 45, NGC 2244 183, NGC 2244 271 and HD 37357 (from top to bottom) discovered using the MOST satellite. The original amplitude spectra are shown in gray, while the frequencies identified as pulsation modes are marked with black lines. Gray frequencies that seem to be statistically significant, but are not identified as pulsation modes, are alias frequencies with the MOST orbit; these are separated from the pulsation frequencies by the MOST orbital frequency of ~164.4μHz and disappear when the true pulsation frequencies are prewhitened during the data analysis.

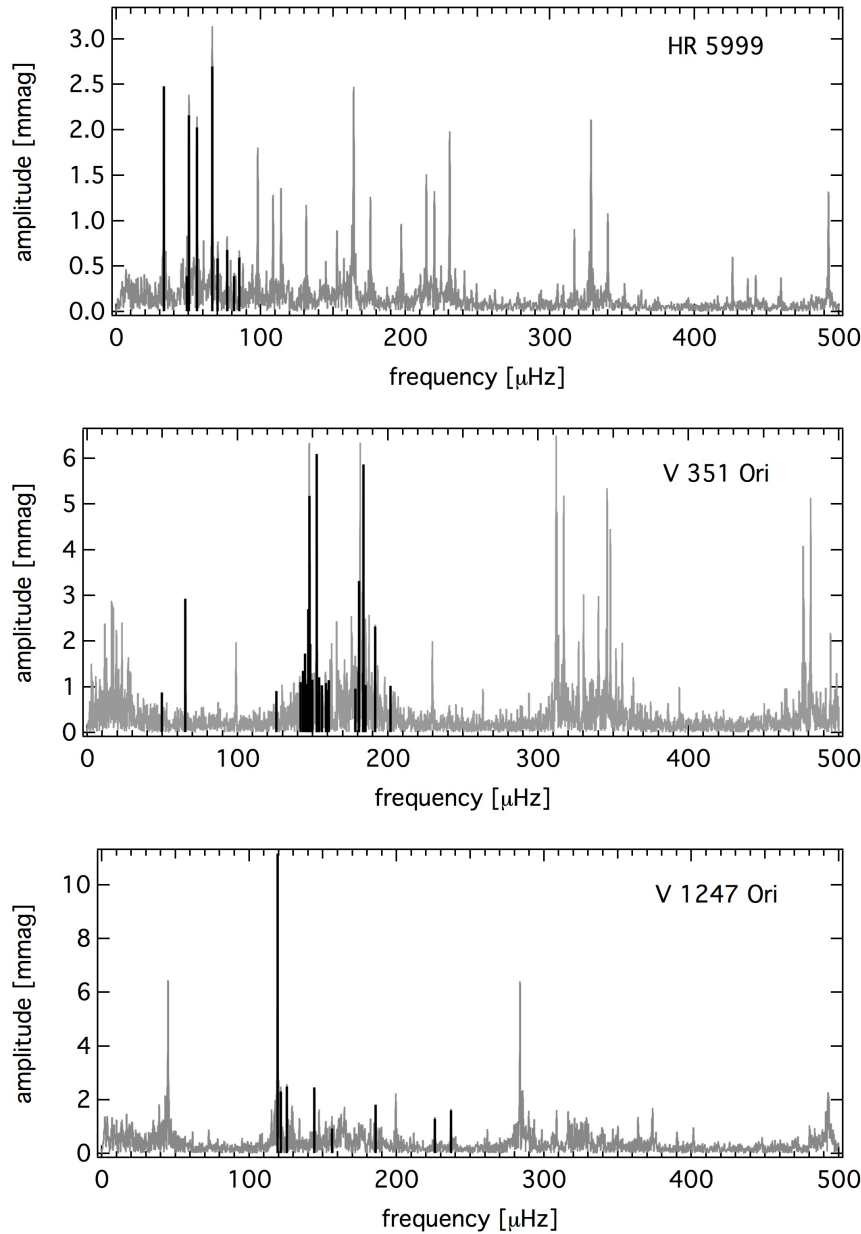

**Fig. S2.** Amplitude spectra for the three pulsating pre-MS stars HR 5999, V 1247 Ori and V 351 Ori obtained from newly acquired data from the MOST space telescope. The original amplitude spectra are shown in gray, while the frequencies identified as pulsation modes are marked with black lines. Grey frequencies that seem to be statistically significant, but are not identified as pulsation modes, are alias frequencies with the MOST orbit; these are separated from the pulsation frequencies by the MOST orbital frequency of ~164.4µHz and disappear when the true pulsation frequencies are prewhitened during the data analysis.

## 1.1. MOST observations

Throughout the ten years of its operation, the MOST space telescope (*9*) has spent dedicated observing time on 12 pre-MS pulsators and candidates, sometimes collecting data a year after the initial observations for a longer time base and higher precision, e.g. for HD 142666 (*24*). MOST also observed the young cluster NGC 2264 for the first time from December 7, 2006, to January 3, 2007, in a dedicated exploratory observing run on the cluster itself (*25*). The second MOST observing run on NGC 2264 lasted from December 5, 2011, to January 14, 2012, and was conducted as part of the *CSI2264* (Coordinated Synoptic Investigation of NGC 2264) project together with the space telescopes CoRoT (*10*), Spitzer (*26*) and Chandra (*27*).

Due to the magnitude range (7 < V < 12 mag) and the large number of targets, NGC 2264 was observed in the open field of the MOST Science CCD in Guide Star Photometry Mode during both runs (*25*). Because not all of the candidate pulsating pre-MS stars could be reached using a single pointing of the satellite, two fields of observations were chosen and observed in alternating halves of each 101-min orbit. Using this setting, MOST time series photometry was obtained for a total of 68 stars in the region of NGC 2264 in the 2006 run (*25*) and 67 in the 2011/12 run.

## 1.2. CoRoT observations

The main research areas of the CoRoT satellite mission (*10*) are the search for extrasolar planets and asteroseismology of main sequence stars. The first observations of the young open cluster NGC 2264 for 23.4 days in March 2008 (i.e., the Short Run SRa01) were conducted in the context of the Additional Program (AP; *28*). A second short run, (i.e., SRa05) on the cluster NGC 2264 was conducted in December 2011 / January 2012 with a time base of about 39 days, as part of the *CSI2264* project described above. For both observing runs, the complete cluster was placed in one Exofield CCD and data were taken for all stars in the accessible magnitude range, i.e., from 10 to 16 mag in R. The 100 brightest stars in the field of NGC 2264 were the primary targets to search for stellar pulsations among pre-MS cluster members.

The reduced data were extracted from the CoRoT data archive. The CoRoT data reduction pipeline (*29*) flags those data points that were obtained during passages of the satellite over the South Atlantic Anomaly and replaces them with linearly interpolated values. We did not use these interpolated data points in our analysis.

## 1.3. Frequency analyses

For the frequency analyses, we used the software package Period04 (*30*) that combines Fourier and least squares analyses. Frequencies were prewhitened and considered to be significant if their amplitudes exceeded four times the local noise level in the amplitude spectrum (*31*, *32*).

We verified the analysis using the SigSpec software (*33*). SigSpec computes significance levels for amplitude spectra of time series with arbitrary time sampling. The probability density function of a given amplitude level is solved analytically and the solution includes dependencies on the frequency and phase of the signal.

Only frequencies that were found using both methods and not related to any instrumental effects (e.g., caused by the satellites' revolutions in their low earth orbits) were considered to be intrinsic to the stars.

## 2. High-resolution spectroscopy: observations, data reduction and analysis

### 2.1. Observations and data reduction

High-resolution spectroscopic measurements with a signal-to-noise ratio above 100 have been obtained from the Mc Donald 2.7m telescope with the Tull échelle spectrograph, the ESO 3.6m telescope with HARPS, the CFHT with ESPaDOnS, the 1.2m Mercator telescope with HERMES, and the low-resolution R-C spectrograph at the CTIO 1.5m telescope. For two stars, the spectroscopic data available in the ESO HARPS archive and for one star in the ESO UVES archive were included in our analysis. An overview is given in Table S1.

In the adopted configuration, the cross-dispersed Robert G. Tull échelle spectrograph at the McDonald 2.7m telescope has a resolving power of 60 000. Each spectrum covers the wavelength range from 3633 – 10849Å with gaps between the échelle orders at wavelengths longer than 5880Å. Bias and Flat Field frames were obtained at the beginning of each night, while several Th-Ar comparison lamp spectra were obtained each night for wavelength calibration purposes. The reduction was performed using the Image Reduction and Analysis Facility IRAF (*http://iraf.noao.edu*).

The data obtained with the fibre-fed high-resolution échelle spectrograph HARPS (*34*) at the ESO La Silla 3.6m telescope were either collected as CoRoT or complementary targets of the ESO LP185.D-0056 or taken from the online ESO HARPS archive. In the adopted EGGS configuration HARPS has a resolving power of 80 000. Each spectrum covers the wavelength range from 3781 – 6913Å, and, hence, includes the hydrogen Balmer lines H$\alpha$, H$\beta$, H$\gamma$, H$\delta$ and H$\varepsilon$. The spectra were reduced using the ESO and a semi-automatic MIDAS pipeline (*35*).

The cross-dispersed échelle spectrograph UVES is mounted at the ESO Very Large Telescope (VLT) and yields a resolving power of 40 000 in the standard mode. All reduction steps were performed within the UVES pipeline (version 5.2.0) and the Reflex software (http://www.eso.org/sci/software/reflex).

The ESPaDOnS spectropolarimeter of the Canada-France-Hawaii Telescope (CFHT) consists of a table-top, cross-dispersed échelle spectrograph fed by a double optical fiber directly from a Cassegrain mounted polarization analysis module. Stokes I and V spectra were obtained throughout the 3700–10400Å spectral range at a resolving power of about 65 000. The spectra were reduced using the Libre-ESpRIT reduction pipeline (*36*).

For the star V 351 Ori we obtained a high-resolution spectrum in March 2014 with the HERMES fiber-fed spectrograph (*37*) at the 1.2m Mercator telescope on La Palma. In its high resolution mode, HERMES has a resolving power of about 85 000. Data were reduced using the automated pipeline and radial velocity toolkit HermesDRS.

Classification spectra of low resolution (R ~ 3000) in the wavelength range from 4050 to 4700Å were obtained for four stars of the present sample with the CTIO 1.5m telescope and R-C spectrograph. Data reduction (*38*) was performed using the IRAF (*http://iraf.noao.edu*) software package.

## 2.2. Spectrum analysis

The spectroscopic analysis was performed using the LLMODELS model atmosphere code (*39*), the VALD database for atomic line parameters (*40*), SYNTH3 (*41*) for the computation of synthetic spectra and an updated version of the SME (Spectroscopy Made Easy) software package (*42*). Using these methods for each of the 34 stars, the effective temperature ($T_{eff}$), surface gravity (log $g$), projected rotational velocity ($v \sin i$) and metallicity ($[Fe/H]$) were determined. For all stars for which a metallicity ([Fe/H]) value has not already been given in the literature, we determined the metallicity by fitting the iron abundance of the least blended iron lines.

For the four stars observed with the CTIO 1.5m telescope we only have low-resolution spectra, which did not allow us to determine [Fe/H] values. Additionally, the error bars on $T_{eff}$, log $g$ and $v \sin i$ are significantly larger compared to the values determined from the other spectroscopic observations.

The critical breakup velocity was computed for all 34 stars according to the formula

$$v_{crit} = \sqrt{\frac{2}{3}\frac{GM}{R}}$$

where $G$ is the gravitational constant, $M$ is the stellar mass, $R$ the stellar radius and the factor 2/3 accounts for the non-sphericity of the star. $M$ and $R$ were derived from the pre-MS evolutionary tracks. The histogram of the percentage of $v\sin i / v_{crit}$ shows that all the pre-MS stars of our sample show rotation rates lower than 50% of their breakup velocity.

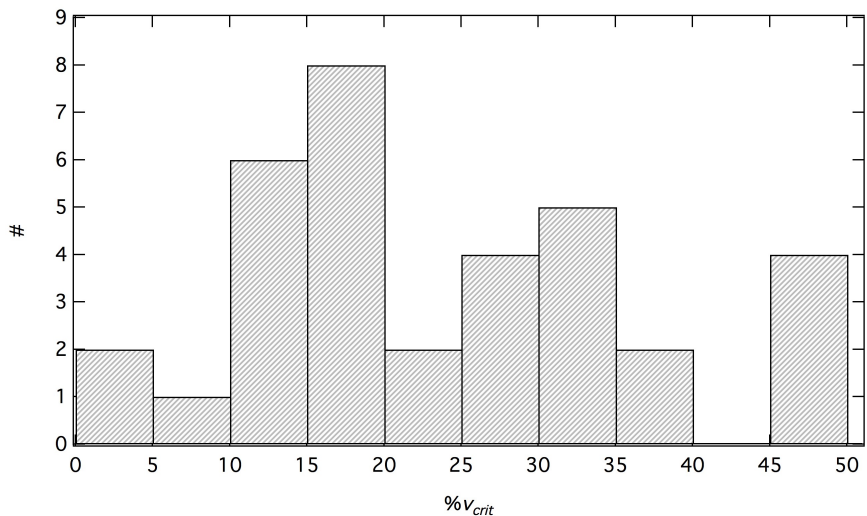

**Fig. S3.** Histogram of the projected rotational velocity, expressed as percentage of critical breakup velocity, $v_{crit}$, measured for the 34 pulsating pre-MS stars in our sample.

## 3. Rotation versus pulsation as cause of $f_{max}$

Young stars are known to show variability caused by inhomogeneities in a circumstellar shell or disk, or variations caused by magnetic effects (e.g., star spots) in the stellar photosphere. Therefore, we investigated whether the seismic quantity upon which we rely in this work, $f_{max}$, could be explained by rotational modulation rather than by stellar pulsation.

The time scale of variations in a circumstellar shell will affect mostly the low-frequency region of the spectra. Sporadic, rapid effects due to interaction with the shell do not persist in the frequency spectra. Most of the stars with accretion bursts, for example, do not show a significant periodic signal. In their study of young stellar objects in NGC 2264, Stauffer et al. (*43*) report that only four stars have strong periodicities, which lie between 6 and 7 days, corresponding to frequencies between 1.938 and 1.653 µHz. This is much lower than the frequencies between ~46 and ~1020 µHz we attributed to pulsation. Moreover, in the presence of both pulsations and rotation, it is mostly possible to discriminate the frequencies from rotation and pulsation in an additional way in the multiperiodic signals occurring in Fourier space, because the rotational frequency will often occur with multiples, while this is not the case for the pulsational frequencies in the linear regime (see e.g., *44, 45*). All frequencies identified by us as being caused by pulsations do not show multiples in the amplitude spectra and are all independent from each other (i.e., there are also no linear combinations present). In particular, the highest observed pressure-mode frequency, $f_{max}$, is in none of the cases a multiple of any other frequency or a linear combination of several frequencies.

We conclude that the multiperiodicity detected for the pre-MS stars in our sample is of such nature that the values of $f_{max}$ are not due to the rotation of the stars.

## 4. Pre-MS stars in the evolutionary diagrams

Pre-MS tracks were computed using the YREC evolution code (*46*) with physics as described in Guenther et al. 2009 (*12*) adopting solar metallicity (i.e., Z = 0.02). The masses and relative ages listed in Table S1 are taken from the pre-MS evolutionary tracks. Relative ages are Kelvin-Helmholtz time scales (*14*) and are set to zero on the Hayashi track. As any change to the structure of the environment of the young star will affect the Kelvin-Helmholtz time scale, the uncertainties in these ages are rather high. Hence, we consider them only as relative ages to illustrate the pre-MS time scale of evolution at different masses.

The exact location of the pre-MS evolutionary tracks strongly depends on the initial conditions and on the input physics, where the metallicity is one of the most influential factors. In the present study we adopted solar metallicity for the evolutionary tracks as a good average value for our sample of stars (see Table S1).

The locations of all 34 pre-MS pulsators in the evolutionary diagram are shown in Fig. S4. The trend clearly seen using only the members of NGC 2264 is confirmed for this more diverse sample. The two red symbols around log $T_{eff}$ of 3.87 and log $g$ of 4.0 are two stars in the field of view of the cluster NGC 2244 (i.e., NGC 2244 183 and NGC 2244 399) for which the membership is very uncertain *47, 48, 49*). If NGC 2244 183 and NGC 2244 399 are indeed non-members of the cluster, they would be field stars of hitherto unknown evolutionary stage. In that case, we propose that their non pre-MS nature is supported by their pulsational properties in the evolutionary diagram.

The relation between the seismic properties and the relative stage in the pre-MS evolution can also be seen in Fig. S5, where $f_{max}$ is drawn together with radius (R/R$_\odot$) and relative age in million years (Myr).

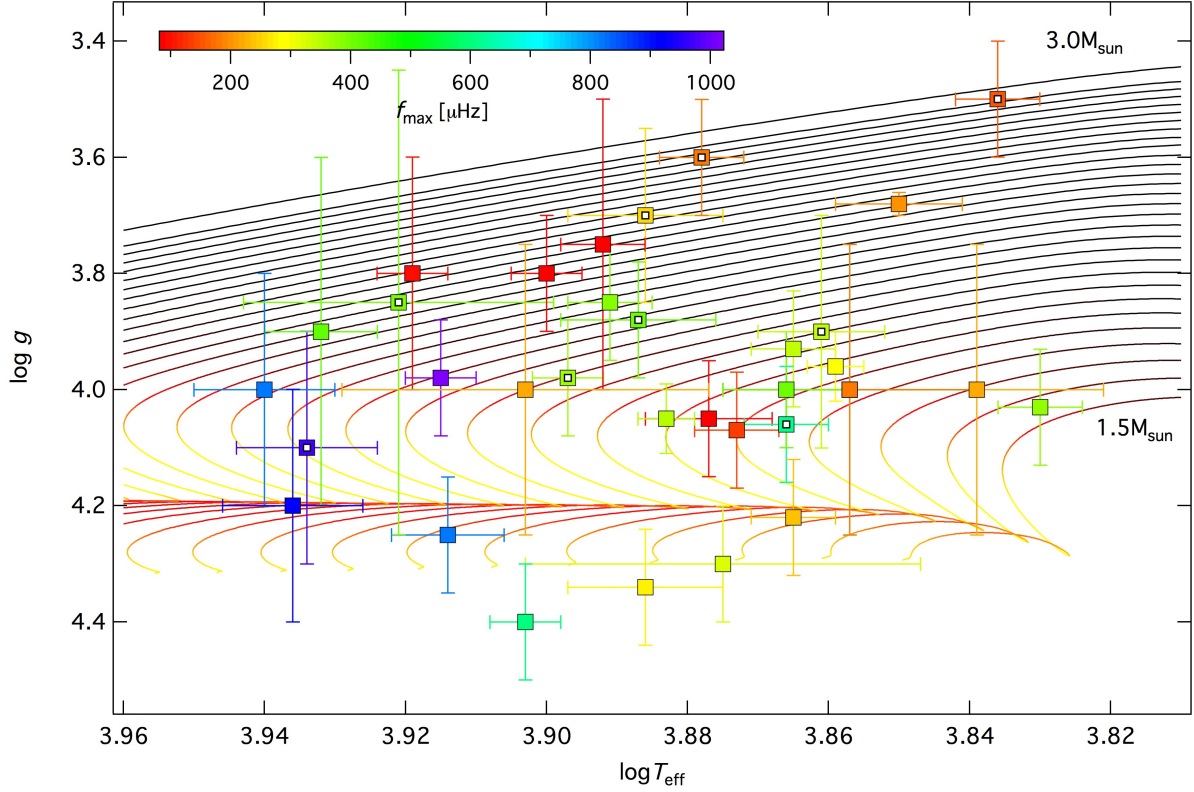

**Fig. S4.** Evolutionary diagram of surface gravity against effective temperature showing the 34 pulsating pre-MS stars with $f_{max}$ represented by color (according to the bar at top-left). Pre-MS evolutionary tracks (*12*) are shown from 1.5 to 3.0 M$_\odot$ in steps of 0.05 M$_\odot$ and are color-coded according to the contribution of gravitational potential energy to the total energy output: when gravitational contraction is the only energy source, the pre-MS tracks are drawn in black; the less efficient gravitational contraction becomes closer to the ZAMS due to the onset of the first nuclear burnings that do not yet occur in equilibrium, the brighter the pre-MS tracks become. The symbols marked with a white smaller square in their centers are members of the young cluster NGC 2264 and are labeled "a" to "i" in Fig. 2 and Fig. 3.

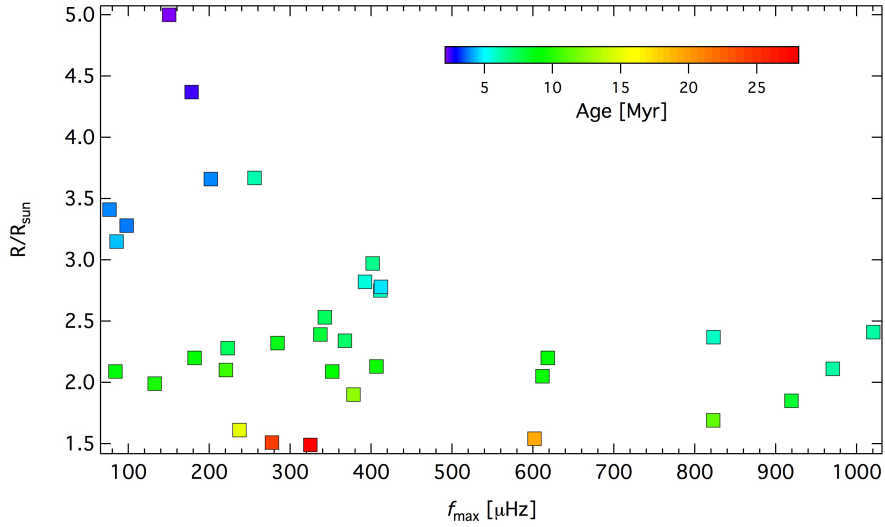

**Fig. S5.** Relation of the maximum pulsation frequency, $f_{max}$, to radius and age (shown by color according to the scale at top right). The lowest $f_{max}$ values are found in the youngest objects with the largest radii (blue symbols); with increasing age, the stellar radii shrink and $f_{max}$ increases.

In Fig. S6, we illustrate how the highest significant observed pressure-mode frequency, $f_{max}$, which was used as a proxy for the acoustic cut-off frequency, compares to the value of this quantity computed from current seismic models (*12*). This comparison can be used for future improvements of the theoretical pre-MS models in terms of the assumed input physics, keeping in mind that the observed $f_{max}$ is by definition a lower limit of the true acoustic cut-off frequency.

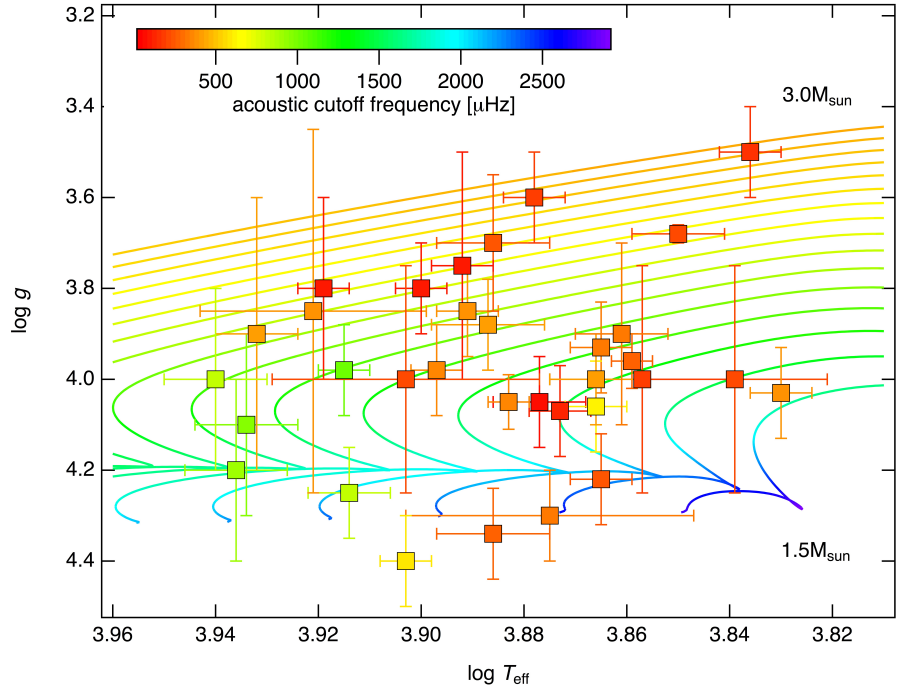

**Fig. S6.** Theoretical prediction for the acoustic cutoff-frequency (solid lines; shown by color according to the scale in the top left, *12*) in the evolutionary diagram and comparison to the maximum observed pulsation frequency, $f_{max}$ (squares; shown by color according to the scale in the top left).

| Star | $f_{max}$ | $f_{max}$ data | $f_{max}$ Ref | $\log T_{eff}$ | $\log g$ | $v\sin i$ | [Fe/H] | $M$ | %$v_{crit}$ | age | source |
|---|---|---|---|---|---|---|---|---|---|---|---|
| | (μHz) | | | (dex) | (dex) | (kms$^{-1}$) | (dex) | (M$_\odot$) | (%) | (Myr) | |
| β Pic | 822.454 | gb | *50* | 3.914±0.008 | 4.25±0.10 | 124±7 | -0.02 | 1.8 | 32.8 | 11 | *51* |
| EE Cha | 392.130 | gb | *52* | 3.891±0.006 | 3.85±0.10 | 85±5 | -0.03 | 2.1 | 22.2 | 6 | H |
| EF Cha | 336.921 | gb | *52* | 3.865±0.006 | 3.93±0.10 | 68±4 | -0.18 | 1.8 | 18.0 | 8 | H |
| HD 104237 | 412.037 | gb | *52* | 3.932±0.008 | 3.90±0.30 | 10±1 | +0.16 | 2.3 | 2.6 | 5 | *53* |
| HD 142666 | 324.653 | M | *54* | 3.875±0.028 | 4.30±0.10 | 66±4 | +0.20 | 1.6 | 17.7 | 28 | *55* |
| HD 144277 | 822.789 | M | *56* | 3.940±0.010 | 4.00±0.20 | 60±4 | -0.09 | 2.1 | 15.6 | 6 | H |
| HD 34282 | 919.248 | M | *57* | 3.936±0.010 | 4.20±0.20 | 129±8 | -0.80 | 2.0 | 33.7 | 8 | *58* |
| HD 37357 | 1020.336 | M | *new* | 3.915±0.005 | 3.98±0.10 | 180±11 | -0.10 | 2.0 | 47.1 | 6 | T |
| HR 5999 | 85.078 | M | *21 & new* | 3.892±0.006 | 3.75±0.25 | 120±7 | -0.12 | 2.4 | 30.8 | 4 | U* |
| IC348 H254 | 85.718 | gb | *59* | 3.900±0.005 | 3.80±0.10 | 107±6 | -0.18 | 2.3 | 27.6 | 4 | T |
| IP Per | 601.852 | gb | *60* | 3.903±0.005 | 4.40±0.10 | 70±4 | -0.24 | 1.7 | 18.6 | 20 | T |
| NGC 2244 183 | 132.176 | M | *new* | 3.873±0.006 | 4.07±0.10 | 174±10 | +0.07 | 1.7 | 46.2 | 10 | T |
| NGC 2244 271 | 277.315 | M | *new* | 3.886±0.011 | 4.34±0.10 | 52±3 | -0.07 | 1.6 | 13.8 | 24 | T |
| NGC 2244 399 | 83.657 | M | *new* | 3.877±0.009 | 4.05±0.10 | 172±10 | -0.19 | 1.7 | 45.8 | 9 | T |

| Star | $f_{max}$ | $f_{max}$ data | $f_{max}$ Ref | $T_{eff}$ | $\log g$ | $v\sin i$ | [Fe/H] | M | Age | Ref | Notes |
|---|---|---|---|---|---|---|---|---|---|---|---|
| NGC 2244 45 | 97.778 | M | new | 3.919±0.005 | 3.80±0.10 | 150±9 | -0.52 | 2.4 | 38.5 | 4 | T |
| NGC 6383 55 | 220.185 | gb | 61 | 3.839±0.018 | 4.00±0.50 | <50 | 0.00 | 1.6 | 6.7 | 11 | RC |
| NGC 6530 159 | 618.252 | gb | 62 | 3.857±0.018 | 4.00±0.50 | <50 | 0.00 | 1.7 | 13.3 | 9 | RC |
| NGC 6530 38 | 181.528 | gb | 63 | 3.857±0.016 | 4.00±0.50 | 100±25 | 0.00 | 1.7 | 26.6 | 9 | RC |
| NGC 6530 57 | 222.488 | gb | 63 | 3.903±0.026 | 4.00±0.50 | <50 | 0.00 | 1.9 | 13.1 | 7 | RC |
| PDS 2 | 377.778 | gb | 64 | 3.830±0.006 | 4.03±0.10 | 17±1 | -0.18 | 1.5 | 4.4 | 13 | H* |
| RS Cha a | 351.620 | gb | 65 | 3.883±0.004 | 4.05±0.06 | 64±4 | +0.17 | 1.7 | 17.0 | 9 | 65 |
| RS Cha b | 284.259 | gb | 65 | 3.859±0.004 | 3.96±0.06 | 70±4 | +0.17 | 1.7 | 18.6 | 9 | 65 |
| V 1247 Ori | 236.956 | M | 22 & new | 3.865±0.006 | 4.22±0.10 | 66±4 | -0.33 | 1.6 | 17.6 | 15 | H* |
| V 346 Ori | 406.331 | gb | 66 | 3.866±0.009 | 4.00±0.10 | 125±8 | -0.05 | 1.7 | 33.2 | 9 | 66,67 |
| V 351 Ori | 201.720 | M | 23 & new | 3.848±0.009 | 3.68±0.20 | 102±6 | -0.16 | 2.3 | 26.0 | 4 | He |
| HD 261230 (c) | 243.403 | M | 25 & new | 3.886±0.011 | 3.70±0.15 | 95±6 | -0.17 | 2.0 | 24.9 | 6 | E, T |
| HD 261387 (f) | 401.690 | M | 25 & new | 3.921±0.010 | 3.83±0.20 | 140±8 | +0.04 | 1.9 | 36.7 | 7 | E, T |
| HD 261711 (i) | 970.370 | C | 68 | 3.934±0.010 | 4.10±0.20 | 53±3 | -0.15 | 2.0 | 13.8 | 6 | T |
| HD 261737 (e) | 465.451 | C | new | 3.897±0.005 | 3.98±0.10 | 191±11 | -0.04 | 1.9 | 50.1 | 7 | T |
| HD 262014 (g) | 485.197 | C | new | 3.887±0.011 | 3.88±0.10 | 51±3 | -0.09 | 2.0 | 13.3 | 6 | T |
| HD 262209 (d) | 342.370 | C | new | 3.861±0.009 | 3.90±0.20 | 68±4 | -0.06 | 1.7 | 29.6 | 7 | T |
| NGC 2264 104 (h) | 646.933 | C | 25 & new | 3.866±0.006 | 4.06±0.10 | 106±6 | -0.09 | 1.7 | 28.1 | 10 | T |
| V 588 Mon (b) | 177.778 | C | 8 | 3.878±0.006 | 3.60±0.20 | 132±8 | +0.10 | 2.8 | 33.3 | 3 | E, H, T |
| V 589 Mon (a) | 152.176 | C | 8 | 3.836±0.006 | 3.50±0.10 | 54±3 | -0.08 | 2.9 | 13.7 | 2 | E, H, T |

**Table S1**: Photometric and spectroscopic parameters for the 34 pulsating pre-MS stars. The nine members of NGC 2264 are listed at the end of the table with the labels "a" to "i" used in the main text next to their identifiers. The columns contain star name, the highest observed pressure-mode pulsation frequency ($f_{max}$), reference to the data used to determine $f_{max}$ ($f_{max}$ data) – MOST (M) data, CoRoT (C) data, or ground-based observations (gb), references to publications if the data are published already or marked as "*new*" if previously unpublished data were used and are shown in Fig. S1 and Fig. S2 ($f_{max}$ Ref), effective temperature ($T_{eff}$), surface gravity ($\log g$), projected rotational velocity ($v\sin i$) and metallicity ([Fe/H]). Masses (M) and relative ages are read from the pre-MS evolutionary tracks. The masses are used to compute the percentage of critical velocity (%$v_{\rm crit}$). The last column lists the source for the spectroscopic parameters: own spectrum analysis using data from McDonald Observatory Tull spectrograph (T), ESO HARPS (H) or HARPS archive (H*), ESO UVES archive (U*), CFHT ESPaDOnS (E), Mercator telescope with HERMES (He) and the CTIO R-C spectrograph (RC) or references when data were taken from the literature. The low-resolution spectra for the four stars observed at CTIO (RC) did not allow us to determine the [Fe/H] values, and, hence we assumed solar metallicity. Note that the error bars in $T_{\rm eff}$ and log$g$ do not include the uncertainties of the input physics of the underlying atmospheric models, and, hence can only be considered as lower limits.